\documentclass[12pt]{article}
\usepackage{axodraw}
\usepackage{amsmath}
\usepackage{graphicx}

\def\be{\begin{equation}}
\def\ee{\end{equation}}
\def\bea{\begin{eqnarray}}
\def\eea{\end{eqnarray}}

\def\lsim{\raise0.3ex\hbox{$\;<$\kern-0.75em\raise-1.1ex\hbox{$\sim\;$}}}
\def\gsim{\raise0.3ex\hbox{$\;>$\kern-0.75em\raise-1.1ex\hbox{$\sim\;$}}}

\def\inbar{\,\vrule height1.5ex width.4pt depth0pt}

\def\IC{\relax\hbox{$\inbar\kern-.3em{\rm C}$}}
\def\IQ{\relax\hbox{$\inbar\kern-.3em{\rm Q}$}}
\def\IR{\relax{\rm I\kern-.18em R}}
 \font\cmss=cmss10 \font\cmsss=cmss10 at 7pt
\def\IZ{\relax\ifmmode\mathchoice
 {\hbox{\cmss Z\kern-.4em Z}}{\hbox{\cmss Z\kern-.4em Z}}
 {\lower.9pt\hbox{\cmsss Z\kern-.4em Z}}
 {\lower1.2pt\hbox{\cmsss Z\kern-.4em Z}}\else{\cmss Z\kern-.4em Z}\fi}

\begin{document}



\rightline{LPT--Orsay 09/59}

\begin{center}

{\Large {\bf
A clear Dark Matter gamma ray line generated by the Green-Schwarz mechanism
}}
\vspace{1 cm}\\

{\large  Y. Mambrini$^1$
}
\vspace{1cm}\\

$^1$
Laboratoire de Physique Th\'eorique,
Universit\'e Paris-Sud, F-91405 Orsay, France
\vspace{0.3cm}\\

\end{center}

\vspace{1cm}

\abstract{\noindent
We study the phenomenology of a $U_X(1)$ extension of the Standard Model where the SM particles
are not charged under the new abelian group.
The Green-Schwarz mechanism insures that the model is anomaly free.
The erstwhile invisible dark gauge field $X$, even if
 produced with difficulty at the LHC has however a clear signature in gamma-ray telescopes.
 We investigate what BSM scale (which can be interpreted as a low-energy string scale)
  would be reachable by the FERMI/GLAST telescope after
 5 years of running and show that a 2 TeV scale can be testable, which is highly competitive
 with the LHC.
}

\newpage

\vspace{3cm}

\newpage


\vspace{3cm}


\pagestyle{plain}

\section{Introduction}

Recent experiments have shown that dark matter (DM) makes up about $25 \%$ of
the Universe's energy budget, but
its nature is not yet understood \cite{evidence1,evidence2,evidence3,evidence4}.
A natural candidate is a class of weakly coupled massive particle (WIMP). Such a
100 GeV candidate would naturally give the right order of magnitude for the thermal
relic abundance.
One of the most studied extensions of the Standard Model (SM) is the Minimal Supersymmetric Model MSSM
(mSUGRA in its local version) which extends the matter spectrum
thanks to an extension of the space-time
symmetries (see \cite{SUSYDM} and references therein).
Alternative ideas have included attempts to study the dark matter consequences of
the simplest gauge extension of the SM by adding a hidden sector charged under a
new $U_X(1)$ symmetry  \cite{zprime, Feng} and some
specific signatures in gamma ray telescope were discussed. In
\cite{Stuck} it was shown that the Stueckelberg $Z'$ extension of the SM leads to a
viable milli-charged dark matter candidate, whereas the authors of \cite{Arkani}
invoke a multiplet of states and an extra non-abelian
light gauge boson to explain recent experimental anomalies.  In several works \cite{Arkani,Kinetic},
kinematical mixing is necessary to couple the dark sector to the visible one, and
many of these models have exotic
matter, including non-vectorlike matter that couples to both SM and hidden sector gauge group,
the lighter of which would be the DM candidate.
However, none of these works focussed on the consequences of the anomalies induced by these kind of
spectra on DM detection. Recently, some works have studied extensively the LHC prospect of
such a construction \cite{KumarWells,Fucito,Wellsteam} whereas one dimension 6 effective operator approach
leads to specific astrophysical signals \cite{Us}.
In this work, we show that, even if the SM particles are not charged under the extra $U_X(1)$, anomalies
generated by the heavy fermionic spectrum can generate through the Green-Schwarz mechanism,
$XZ\gamma$ effective couplings. This coupling induces a clear $\gamma$ ray line signal observable by GLAST
at energy

\begin{equation}
E_\gamma= m_{DM} \left( 1- \frac{M_Z^2}{4 m_{DM}^2} \right),
\end{equation}

\noindent
where $m_{DM}$ is the WIMP mass. Line emission provides a feature that helps to discriminate against the background.
The article is laid out as follows: we first review the motivations and construction of the effective vertex
that couples the hidden sector $X$ gauge boson to the visible sector. After this,
we compute the relic density generated by the lightest hidden fermion\footnote{which is our natural dark matter candidate whose stability can be ensure by the $U_X(1)$ symmetry.} $\psi_{DM}$ and discuss the
detection of the astrophysical signal and the related uncertainties, before concluding and
comparing with LHC perspective and the predictions of different models.



\section{Effective description of the model}

It is well known that any extension of the SM which introduces chiral fermions
with respect to gauge fields suffers from anomalies, a phenomenon of breaking of gauge
symmetries of the classical theory at one-loop level. Anomalies are responsible for instance for
a violation of unitarity and make a theory inconsistent \cite{ABJ,Coriano}.
For this reason if any construction introduces a new fermionic sector to address
the DM issue of the SM, it is vital to check the cancelation of anomalies
and its consequences on the Lagrangian and couplings.
In this letter, we concentrate on the Green-Schwarz mechanism  which arises automatically
in string theory settings. The idea is to add to the Lagrangian local gauge non-invariant
terms in the effective action whose gauge variations cancel the anomalous triangle diagrams.
There exist two kinds of term which can cancel the mixed $U_X(1)\times G_A^{\mathrm{SM}}$
anomalies, with $U_X(1)$ being the hidden sector gauge group and $G_A^{\mathrm{SM}}$ one
of the SM gauge group $SU(3)\times SU(2)\times U_Y(1)$ :
the Chern Simons (CS) term which couples the $G_A^{\mathrm{SM}}$ to the $U_X(1)$ gauge boson,
and the Peccei-Quinn (PQ, or Wess-Zumino (WZ)) term which couples the $G_A^{\mathrm{SM}}$
gauge boson to an axion. In the effective action, these terms are sometimes called
Generalized Chern--Simons (GCS) terms \cite{Abdk}.
In order to describe the relevant structure,
we can separate the effective Lagrangian into a sum of classically gauge variant and gauge invariant
terms\footnote{We will consider $U_X(1) \times U_Y^2(1)$ mixed anomalies throughout the
paper to simplify the formulae, the generalization to $U_X(1) \times SU^2(N)$ being straightforward.}
\cite{KumarWells,Wellsteam,Abdk} :

\begin{eqnarray}
{\cal L}_{inv} &&= - \frac{1}{4 g'^2} F^{Y \mu \nu} F^Y_{\mu \nu}
- \frac{1}{4 g_X^2} F^{X \mu \nu} F^X_{\mu \nu}
-\frac{1}{2} (\partial_\mu a_X - M_X X_\mu)^2
-i \overline{\psi} \gamma^\mu D_\mu \psi
\nonumber
\\
{\cal L}_{var} &&=
\frac{C}{24 \pi^2} a_X \epsilon^{\mu \nu \rho \sigma} F^Y_{\mu \nu} F^Y_{\rho \sigma}
+\frac{E}{24 \pi^2} \epsilon^{\mu \nu \rho \sigma} X_{\mu} Y_{\nu} F^Y_{\rho \sigma}
\label{Lagrangian}.
\end{eqnarray}

\noindent
The Stueckelberg axion $a_X$ ensures the gauge invariance of the effective Lagrangian and
$g_X$ and $F^X_{\mu \nu}=\partial_{\mu} X_\nu - \partial_\nu X_\mu$
are the gauge coupling and field strength of $U_X(1)$. The axion
has a shift transformation under $U_X(1)$
\begin{equation}
\delta X_{\mu} \ = \ \partial_{\mu} \alpha \quad , \quad \delta a_X
\ = \ \alpha \  M_X .
\end{equation}

\noindent
Notice that our dark matter candidate is expected to be chiral with respect
to the dark sector $U_X(1)$, with a mass of the order of $M_X$
because its mass should be generated by the spontaneous breaking of $U_X(1)$.
This differs considerably from the result obtained with a leptophylic dark sector \cite{Ko}
 which considered vector-like dark matter with a Dirac mass term. A hierarchy
 between $X$ and the lightest heavy fermion charged under $U_X(1)$
 (our natural dark matter candidate) would imply a hierarchy between hidden sector
 Yukawa and gauge couplings. The hidden fermionic sector being chiral,
 looking for the effects of mixed anomalous diagrams on DM phenomenology is not an ad-hoc assumption,
  but a necessity.

\begin{figure}
\begin{center}
\begin{picture}(100,90)(-30,-5)
\hspace*{-11.5cm}
\SetWidth{1.1}
\Text(120,0)[]{\bf a)}
\Photon(120,50)(160,50){3.5}{5}
\Photon(180,40)(210,25){3.5}{5}
\Photon(180,60)(210,75){3.5}{5}
\Line(160,50)(180,60)
\Line(180,60)(180,40)
\Line(160,50)(180,40)
\Text(140,63)[]{$X_{\mu}(p_3)$}
\Text(230,75)[]{$Y_\nu(p_1)$}
\Text(230,25)[]{$Y_\rho(p_2)$}
\Text(290,0)[]{\bf b)}
\Photon(270,50)(300,50){3.5}{5}
\DashLine(300,50)(325,50){3}
\Photon(325,50)(360,25){3.5}{5}
\Photon(325,50)(360,75){3.5}{5}
\Text(285,63)[]{$X_{\mu}(p_3)$}
\Text(315,40)[]{$a_X$}
\Text(380,75)[]{$Y_\nu(p_1)$}
\Text(380,25)[]{$Y_\rho(p_2)$}
\Text(440,0)[]{\bf c)}
\Photon(440,50)(475,50){3.5}{5}
\Photon(475,50)(510,25){3.5}{5}
\Photon(475,50)(510,75){3.5}{5}
\Text(460,63)[]{$X_{\mu}(p_3)$}
\Text(535,75)[]{$Y_\nu(p_1)$}
\Text(535,25)[]{$Y_\rho(p_2)$}
\end{picture}
 \caption{{\footnotesize Three vertices generated respectively by loops diagrams
 ({\bf a}), Peccei-Quinn ({\bf b}) and Chern--Simons ({\bf c}) terms (Eq. \ref{Coupling}).}}
\label{fig:coupling}
\end{center}
\end{figure}
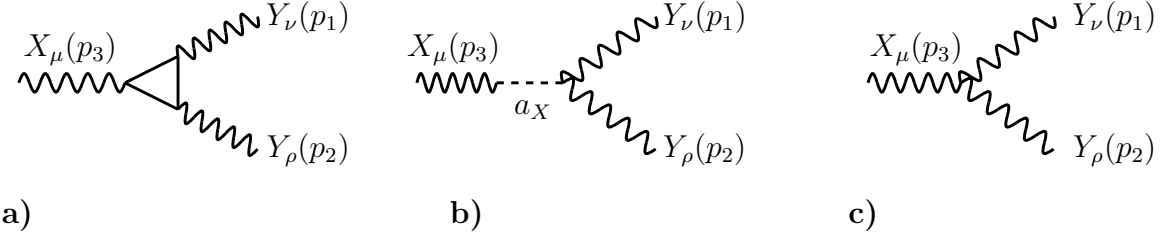

From Eq.(\ref{Lagrangian}) and following \cite{Abdk}, the triangle amplitude
$\left[ X_{\mu}(p_3)Y_{\nu} (p_1) Y_{\rho}(p_2) \right]$ can be
decomposed according to

\begin{eqnarray}
\Gamma_{\mu \nu \rho}^{ZYY}&=&
\left[A_1 ~ \epsilon_{\mu \nu \rho \sigma} p_2^\sigma + A_2 ~ \epsilon_{\mu \nu \rho \sigma} p_1^\sigma
\right.
\nonumber
\\
&+&
\left.
B_1 ~ p_{2\nu} \epsilon_{\mu \rho \sigma \tau } p_2^\sigma p_1^\tau
+ B_2 ~p_{1\nu} \epsilon_{\mu \rho \sigma \tau } p_2^\sigma p_1^\tau
+ B_3 ~p_{2\rho} \epsilon_{\mu \nu \sigma \tau } p_2^\sigma p_1^\tau
+ B_4 ~p_{1\rho} \epsilon_{\mu \nu \sigma \tau } p_2^\sigma p_1^\tau
\right.
\nonumber
\\
&+&
\left.
C~\frac{p_{3\nu}}{p_3^2}\epsilon_{\nu \rho \sigma \tau} p_2^\sigma p_1^\tau
+
E ~\epsilon_{\mu \nu \rho \sigma} (p_2^\sigma-p_1^\sigma)
\right]
\label{Coupling}
\end{eqnarray}

\noindent
where each contribution to the amplitude is as shown in  Fig.\ref{fig:coupling},
and where $B_i$ satisfy

\begin{eqnarray}
B_1(p_1,p_2)&=&-B_4(p_2,p_1)=
-i\frac{g_X g'^2}{8 \pi^2}Tr[Q_X Q_Y Q_Y] I_1
\nonumber
\\
B_2(p_1,p_2)&=&-B_3(p_2,p_1)=
-i\frac{g_X g'^2}{8 \pi^2}Tr[Q_X Q_Y Q_Y] I_2
\end{eqnarray}

\noindent
with $Q_i$ being the $U_i(1)$ charges of the hidden fermions running in the loop,
 and $I_i$  being finite computable integrals (see the appendix of
\cite{Abdk} for details).
Whereas $A_i$ are $UV$ cutoff dependent, following \cite{KumarWells}
the Chern--Simons coefficient can be reabsorbed if we define the coupling
$\tilde A_i=A_i+(-1)^{i+1} E$. The condition for the conservation of the 3
currents gives us 3 Ward identities

\begin{eqnarray}
\tilde A_1 &=& -p_1.p_2 B_1 -p_1^2 B_2
\nonumber
\\
\tilde A_2 &=& -p_1.p_2 B_4 -p_2^2 B_3
\nonumber
\\
C &=& \tilde A_1 - \tilde A_2.
\end{eqnarray}

\noindent
We can now express our effective vertices in terms of  finite integrals:

\begin{eqnarray}
\Gamma_{\mu \nu \rho}^{XZZ}&=&
-2 i \frac{\sin \theta_W^2 g_X g'^2 \mathrm{Tr[Q_X Q_Y Q_Y]}}{8 \pi^2 M^2}
\left(
[\tilde I_1 ~ p_1.p_2 + \tilde I_2 ~ p_1^2]\epsilon_{\mu \nu \rho \sigma} p_1^\sigma
-[\tilde I_2 ~p_2^2 + \tilde I_1 ~ p_1.p_2]\epsilon_{\mu \nu \rho \sigma} p_2^\sigma
\right.
\nonumber
\\
&+&
\left.
[\tilde I_1 ~p_{2\nu} + \tilde I_2 ~p_{1\nu}]\epsilon_{\mu\rho\sigma\tau}p_2^\sigma p_1^\tau
-[ \tilde I_1 ~p_{1\rho} + \tilde I_2 ~p_{2 \rho}]\epsilon_{\mu \nu \sigma \tau}
p_2^\sigma p_1^\tau
\nonumber
\right.
\\
&+& \left.\frac{1}{p_3^2} [\tilde I_1 ~ p_2^2 + \tilde I_2 p_1^2 + (\tilde I_1 + \tilde I_2 )p_1.p_2]
p_{3\mu}\epsilon_{\nu\rho\sigma\tau} p_2^\sigma p_1^\tau
\right)
\label{XZZcoupling}
\end{eqnarray}

\begin{eqnarray}
\Gamma_{\mu \nu \rho}^{X Z \gamma}&=&
-2\left( \cos \theta_W / \sin \theta_W \right) \Gamma_{\mu \nu \rho}^{XZZ}
\label{XZgcoupling}
\end{eqnarray}

\begin{equation}
\Gamma^{X \psi_{DM} \psi_{DM}} = \frac{g_X}{4}\gamma^\mu
\left(
[q_X^R + q_X^L] + [q_X^R-q_X^L]\gamma^5
\right)
\label{XDMDMcoupling}
\end{equation}

\noindent
where we defined $\tilde I_i= M^2 I_i$, $\tilde I_i$ being a dimensionless integral
and $M$ the $U_X(1)$ breaking scale (typically the masses of the hidden fermions
running in the loops).  This scale can be thought of as coming from effective derivative couplings
as was explicitly shown in \cite{Wellsteam,Us}.
One might expect a $\gamma \gamma$ decay channel as well, but this decay is
forbidden when both outgoing particles are massless. It is straightforward to generalize
the formulae in the case of extra fermions charged under $SU(2)\times U_Y(1)\times U_X(1)$.
New couplings of the type
$\Gamma_{\mu\nu\rho}^{XW^+W^-}$ would be generated, but would not change the
general conclusion of our study. Indeed, the annihilation channel
$\psi_{DM} \psi_{DM} \rightarrow X \rightarrow W^+ W^-$ would give a spectrum quite similar to the
one generated by the annihilation into ZZ \cite{Us}. We made the analysis in a specific
"next to minimal" $SU(2)\times U_Y(1)\times U_X(1)$ extension and found that generically
the perspective of detection are even larger in such scenario than in the minimal
$U_X(1)$ extension studied here. Our results are quite conservative in that sense.

Another interesting point is that, in order to generate dynamical masses from the
$U_X(1)$ breaking, the hidden fermions (dark matter candidate included) should be chiral
with respect to $U_X(1)$ with Weyl charges $q_X^L$ and $q_X^R$. This situation is quite different
from many other works concerning leptophylic dark matter \cite{Ko}. Indeed, to deal with
anomaly free models, the DM candidate is taken to be Dirac, with the Dirac masses
having no dynamical origin. In our case, the Green--Schwarz mechanism allows for chiral
fermions in the theory. In our specific example, whereas $q_X^L-q_X^R$ would be fixed
by the coupling $S \psi_{DM}^L \psi_{DM}^R$, the value $q_X^L+q_X^R$ can be freely
chosen. Without any lack of generality (the charges can be reabsorbed in the
definition of $g_X$) we take $q_X^R=-2$ and $q_X^L=1$ throughout the analysis.

Anomalies in QFT are of great interest for another reason: they could indirectly probe
high energy physics and possibly even stringy effects. This is because anomalies can be thought of
as both UV and IR effects. They are clearly visible in the limit of low--energy effective field
theory which we expect to match to data, yet their UV character implies that the existence
(and resolution) of these anomalies can often be tied to a stringy origin. This stringy origin
 makes the anomaly an interesting candidate for a stringy signature to be observed at colliders
 or in DM detection experiments. In particular, different string constructions do not have
 $U_X(1)$ couplings
 to SM fermions at tree level \cite{IBM}. With this in mind, studying anomaly generated
 couplings and their consequences on the dark matter relic density is of fundamental importance.
 It is also important to mention that even if the spectrum is constructed so the
 model is anomaly free at the quantum level ($\mathrm{Tr[Q_X Q_Y Q_Y]=0}$), triangle diagrams
 anomalies can contribute in a non trivial way to  the $XYY$ vertices by higher dimensional
 operators (even dimension 4 for two extra $U(1)$) \cite{Wellsteam,Us,Coriano}.


\section{Dark matter phenomenology}

\begin{figure}
\begin{center}
\begin{picture}(100,90)(-30,-5)
\hspace*{-11.5cm}
\SetWidth{1.1}
\ArrowLine(205,25)(240,50)
\ArrowLine(205,75)(240,50)
\Photon(240,50)(275,50){3.5}{5}
\Photon(275,50)(310,25){3.5}{5}
\Text(260,60)[]{$X$}
\Photon(275,50)(310,75){3.5}{5}
\Text(325,75)[]{$Z,Z$}
\Text(325,25)[]{$Z,\gamma$}
\Text(185,75)[]{$\psi_{DM}$}
\Text(185,25)[]{$\psi_{DM}$}
\ArrowLine(380,100)(415,75)
\ArrowLine(380,0)(415,25)
\ArrowLine(415,25)(415,75)
\Photon(415,75)(450,100){3.5}{5}
\Photon(415,25)(450,0){3.5}{5}
\Text(460,100)[]{$X$}
\Text(460,0)[]{$X$}
\Text(370,100)[]{$\psi_{DM}$}
\Text(370,0)[]{$\psi_{DM}$}
\end{picture}
 \caption{{\footnotesize Feynman diagrams contributing to the dark matter annihilation.}}
\label{feynannihilation}
\end{center}
\end{figure}
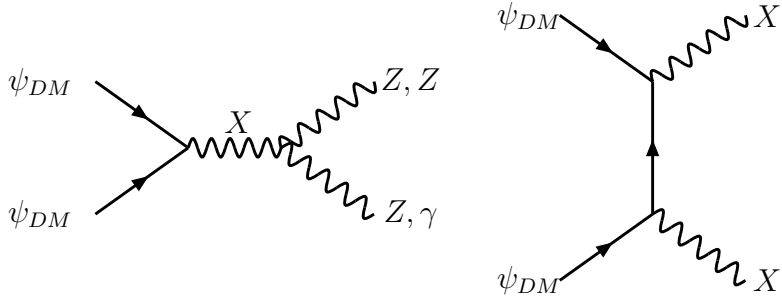

\subsection{Relic abundance}

since $\psi_{DM}$ is the lightest fermionic state that is charged under $U_X(1)$,
the only allowed annihilation final states at tree level are\footnote{
We will suppose that the $X$ boson is lighter than
the chiral matter in the hidden sector. As was emphasized in \cite{KumarWells} this is not an
unreasonable hypothesis as the mass of the most hidden sector matter will be dominated by higher
symmetry breaking scales, and will thus be heavier than $X_\mu$.} $X$ or $Y$.
Concerning the relic abundance of $\psi_{DM}$, we can easily deduce the different annihilation
channels kinematically allowed (Fig.\ref{feynannihilation}) from the effective coupling generated in
Eqs (\ref{XZZcoupling}-\ref{XDMDMcoupling}). Using the symmetries
and current conservation invoked above, one can parameterize the whole model
by 3 physical parameters : $M_X$, $m_{DM}$ and
$\Lambda_X=M \sqrt{8 \pi^2 /Tr[Q_X Q_Y Q_Y] g_X^2 g'^2(\tilde I_2 - \tilde I_1)}$. These low energy parameters
would correspond to $g_X$, $Y_h$ (Yukawa of the lightest fermions in the hidden sector)
and $\langle S \rangle$ (vev of the higgs breaking $U_X(1)$) in a dynamical version of the model.
Our goal is to outline the parameter space that can be probed by the FERMI telescope.

\begin{figure}
    \begin{center}
    \includegraphics[width=2.7in]{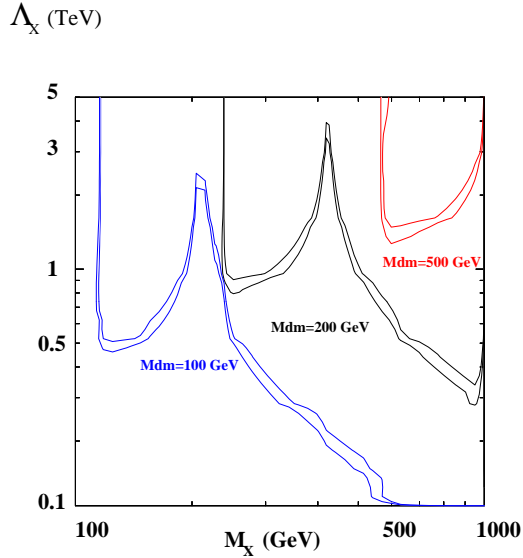}
          \caption{{\footnotesize
Scan on the mass of $X$ (in logarithmic scale) versus the effective scale $\Lambda_{X}$.
Colored lines represent the WMAP limits on the dark matter relic
density for different values of the dark matter mass.
}}
\label{fig:relic}
\end{center}
\end{figure}

We modified the micrOMEGAs code \cite{Micromegas} in order to calculate the relic abundance of
$\psi_{DM}$. It is easy to fulfill the WMAP $5\sigma$ bound \cite{WMAP} on $\mathrm{\Omega_{DM}}$
for a wide range of the DM mass as shown in Fig.\ref{fig:relic}
in the ($M_X, \Lambda_X$) plane for different values of the DM mass (100, 200 and 500 GeV).
We can clearly see the s--channel resonance effect of $\psi_{DM} \psi_{DM} \rightarrow X \rightarrow YY$
 around the region $M_X \simeq 2 m_{DM}$. Within the band, the relic density respects WMAP
constraints whereas below the pole, $\Omega_{DM} < \Omega_{WMAP}$.
We also notice a region where the relic density is below WMAP data
on the left of the plot, where $M_X \lsim m_{DM}$. This region corresponds to
the opening of the t-channel $\psi_{DM} \psi_{DM} \rightarrow X X$ which depends on
$g_X$\footnote{It is important to notice here that
$g_X$ has been absorbed in the definition of $\Lambda_X$ to avoid any explicit dependance
on the coupling. For simplicity, we took $g_X = 2$
throughout the analysis.}. For heavier X ($M_X \gsim 1 $ TeV, $m_{DM} \gsim 500$ GeV)
there still exist regions of the parameter space respecting WMAP. This occurs for
 higher values of $\Lambda_X$ in order to decrease the $XYY$ coupling (proportional to
 $1/\Lambda_X$) to compensate
 the increasing of the relic density proportional to $m_{DM}$.

\subsection{Indirect detection}

The annihilation of WIMP into photons typically proceeds via a complicated set of
processes with an upper cutoff at approximately the WIMP mass.
A line due to the $2\gamma$ or $Z\gamma$ production channel could
be observed as a feature in the astrophysical source spectrum
\cite{LineBuckley}.
In the most popular models, the branching ratio for the annihilation into lines
is typically about $10^{-3}$ or less.
However,
some models exhibit specific signatures as high energy rise due to final state radiation
\cite{Barger,Gammaradiation}.
Other dark matter candidates (like the Inert Higgs Dark Matter
(IHDM)
\cite{IHDM,IHDMline}, extra-dimensional chiral square theories \cite{Chiralsquare}
or neutralino in SUSY models \cite{Neutralino})
can give strong monochromatic
signals from $\gamma \gamma$ or $Z \gamma$ final states.
However, except in \cite{Us}, none of these models exhibit only one $\gamma$ ray line but two or three.
In the Green--Schwarz mechanism, $\gamma \gamma$ final state is excluded by spin conservation,
we are thus left with one monochromatic line, which would be a clear signature of the model.
The spectrum of gamma--rays generated in dark matter annihilations
and coming from a direction forming an angle $\psi$ with respect to
the galactic center is
\begin{equation}
\Phi_{\gamma}(E_{\gamma}, \psi)
=\sum_i
\frac{dN_{\gamma}^i}{dE_{\gamma}}
 Br_i \langle\sigma v\rangle \frac{1}{8 \pi m_{\chi}^2}\int_{line\
 of\ sight} \rho^2
\ dl\ ,
\label{Eq:flux}
\end{equation}
where the discrete sum is over all dark matter annihilation
channels,
$dN_{\gamma}^i/dE_{\gamma}$ is the differential gamma--ray yield,
$\langle\sigma v\rangle$ is the annihilation cross-section averaged
over its velocity distribution, $Br_i$ is the branching ratio of annihilation
into final state ``i'' , and $\rho$ is the dark matter density.

We know that a line due to the $\gamma \gamma$ or Z$\gamma$ channel could be observed in
the astrophysical spectrum \cite{LineBuckley}. Such an observation is a "smoking gun" signal
for WIMP
DM as it is difficult to explain by a process other than WIMP annihilation or decay. Different
models predicts different ratios for such processes. However, in SUSY or KK DM annihilation
the $\gamma$ line is typically loop suppressed as the main annihilation channel contributing
to the thermal relic density dominates and gives a continuous $\gamma$ spectrum. In some of the
parameter space of the IHDM, for DM masses below the $W$ mass, it has been shown in \cite{IHDMline}
that the two $\gamma$ lines could be observed. A similar conclusion was found recently for the chiral square
model \cite{Chiralsquare}. The main reason is that in both models the $Z\gamma$ and the $\gamma \gamma$
are the only kinematically allowed annihilation channel, with a third line being generated by the KK final
 state in \cite{Chiralsquare}.

In our analysis we use diffuse-model simulated data from the centre annulus
(r $\in [20^\mathrm{o}, 35^\mathrm{o}]$),
excluding the region within $15^\mathrm{o}$ of the Galactic plane. It has recently been shown
that it is possible in this case to minimize the contribution
of the Galactic diffuse emission and could give a signal-to-noise ratio up to 12 times greater
than at the Galactic Center (GC)
\cite{Annulusbis}. A very interesting feature of excluding the GC in our analysis
lies in the fact that our results are quite unsensitive to the different dark matter profile
(Einasto \cite{Einasto}, NFW \cite{NFW} or Moore \cite{Moore}) as their contributions
differ largely within the parsec region around the GC.
In addition, we use the LAT line energy sensitivity for 5$\sigma$
detection calculated in \cite{GLASTprelaunch}. The $\gamma$-ray spectrum is calculated
 using an adapted version\footnote{The author wants to thank warmfuly G. Belanger and S. Pukhov for the precious
 help concerning the modification of the code.} of micrOMEGAs \cite{Micromegas}.

\begin{figure}
    \begin{center}
    \hspace{-1cm}
    \includegraphics[width=2.7in]{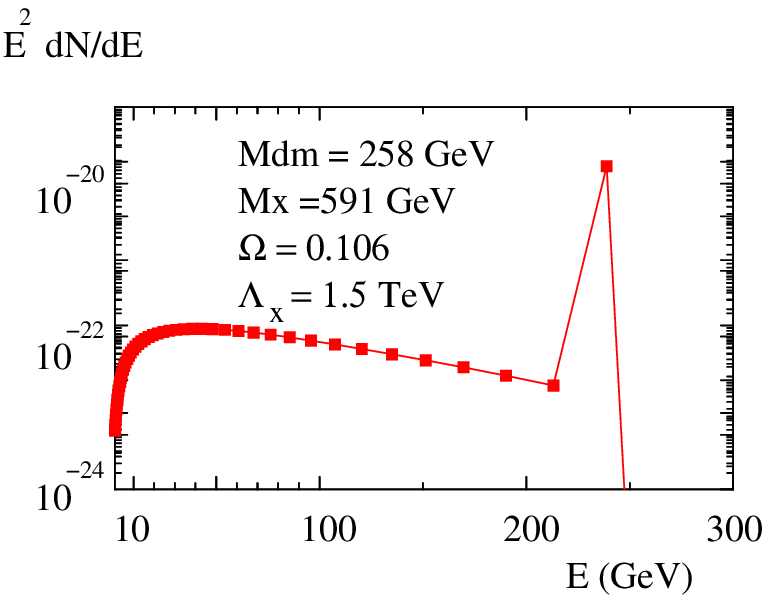}
    \hspace{1cm}
    \includegraphics[width=3.5in]{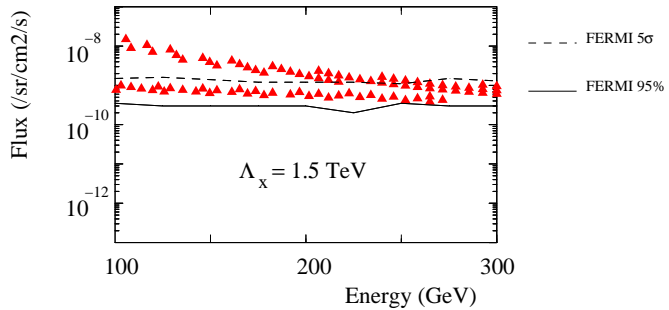}
          \caption{{\footnotesize
Left: example of gamma--ray flux respecting WMAP constraint
for a DM mass of 258 GeV. Right: monochromatic $\gamma-$ray fluxes generated by Green-Schwarz mechanism
in comparison with expected $5\sigma$ and $95\%$ CL sensitivity contours
(5 years of FERMI operation) for the conventional background and unknown WIMP energy,
for an effective scale $\Lambda_X=1.5$ TeV
}}
\label{fig:spectrum}
\end{center}
\end{figure}

\subsection{Results and discussion}

As an illustrative point, we show in the left panel of Fig.\ref{fig:spectrum}
an example of  spectrum from the centre annulus that could be observable by the
FERMI telescope, generated by DM annihilation within the pole region
respecting WMAP constraint(
 $m_{DM} = 258$ GeV and $M_X=591$ GeV).
We can clearly distinguish a $\gamma-$ray line centered
around $E_\gamma= m_{DM} \left( 1- \frac{M_Z^2}{4 m_{DM}^2} \right)$
generated by the s-channel resonance
$\psi_{DM} \psi_{DM} \rightarrow Z \gamma$ above the continuous flux produced by
the annihilation process $\psi_{DM} \psi_{DM} \rightarrow Z Z / Z \gamma$.
We calculated the fluxes generated by the Green-Schwarz mechanism and compared it with the
expected sensitivity of FERMI after 5 years of data-taking \cite{GLAST,GLASTprelaunch}
for 50 energy bins logarithmically distributed between 1 and 300 GeV ($E_k=E_{min} e^{k(E_{max}-E_{min})}$)
and a gamma-ray width of 12$\%$ of $E_{\gamma}$ corresponding to the energy resolution
of FERMI.
The results are presented
in the right panel of Fig.\ref{fig:spectrum}. The Galactic diffuse model used in \cite{GLASTprelaunch}
 is the GALPROP  "conventional" model discussed in \cite{Conventional}. Indeed, the "optimized" model
\cite{Optimized} seems to be disfavored by the non-confirmation of EGRET excess by the
first FERMI released data \cite{GLASTresults}.
We used the results obtained by the full detector Monte Carlo
simulation and reconstruction framework detailed in \cite{GLASTprelaunch}.

\begin{figure}
    \begin{center}
    \hspace{-1cm}
    \includegraphics[width=3.5in]{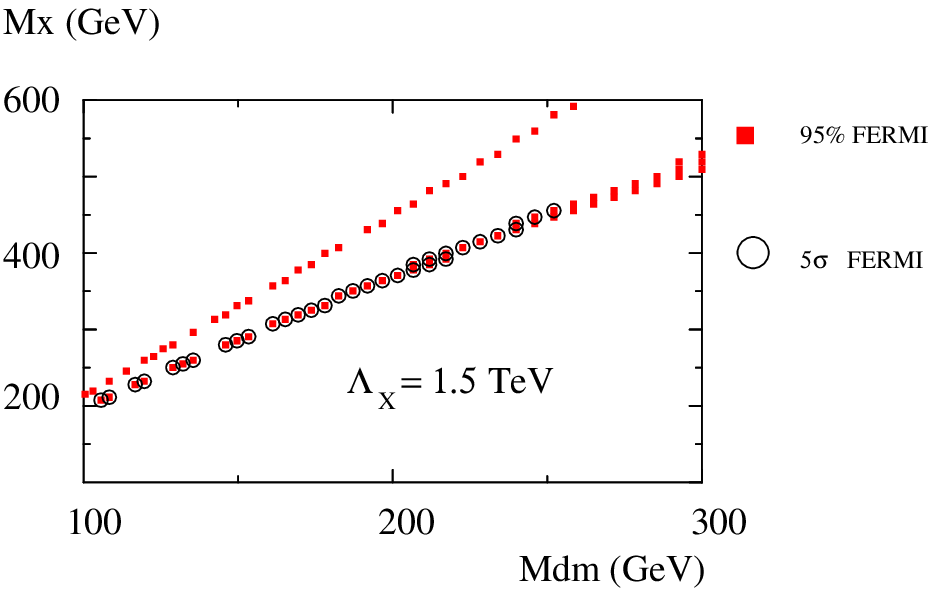}
    \hspace{0.5cm}
    \includegraphics[width=2.9in]{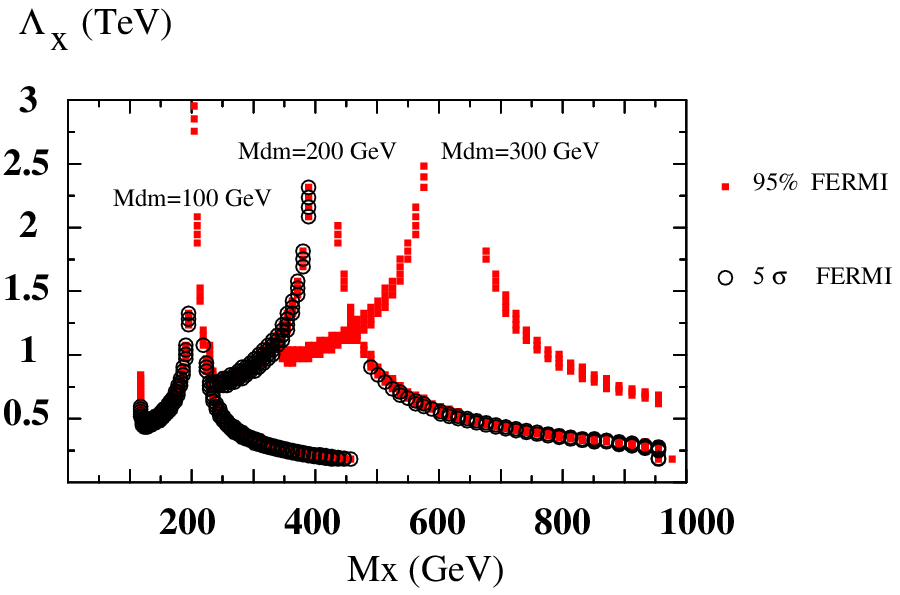}
          \caption{{\footnotesize
Left: scan on $(m_{DM},M_X)$ plane and $\Lambda_X=1.5$ TeV. Points respect WMAP constraints.
Present experiments limits and FERMI sensitivity after 5 years of data taking are displayed.
Right: monochromatic $\gamma-$ray fluxes generated by Green-Scharz mechanism
in comparison with expected $5\sigma$ and $95\%$ CL sensitivity contours
(5 years of FERMI operation) for the conventional background and unknown WIMP energy,
for an effective scale $\Lambda_X=1.5$ TeV
}}
\label{fig:masses}
\end{center}
\end{figure}

We clearly see in the right panel of Fig. \ref{fig:spectrum}
that for $\Lambda_X = 1.5$ TeV, all the parameter space would be observable
by FERMI at 95\% CL. Indeed, the points that respect the WMAP
constraints lie around the p\^ole $M_X \sim 2 m_{DM}$ (see left panel
of Fig.\ref{fig:masses}) where
$\sim 60$\% of the annihilation rate is dominated by the $Z\gamma$ final state.
This proportion still holds for annihilating DM in the Galactic
halo and gives a monochromatic line observable by FERMI. We also show what
values of $\Lambda_X$ are accessible by measurement of
$\gamma-$ray for different values of DM masses (100, 200 and 300 GeV)
in the right panel of Fig. \ref{fig:masses}, where all the points respect the
WMAP relic abundance. We can see that for $m_{DM}\gsim 100$ GeV,
$\Lambda_X \gsim 1$ TeV the model still gives a signal observable by FERMI at 95\% CL.
Obviously, for points lying away from the s-resonance
we still have points which can respect WMAP constraint for lower
values of $\Lambda_X$. All these points (black circles in the right panel of
Fig. \ref{fig:masses}) would be observable at 5$\sigma$ after 5 running years of
FERMI.

Using gamma--ray data from a variety of experiments,
the authors of \cite{Jacques} have calculated conservative
upper limits on the dark matter
annihilation cross section to gamma ray lines over a wide range of masses.
Combining results from COMPTEL, EGRET, HESS and INTEGRAL from different
regions of the sky, and requiring the signal to be as large as the full measured
background in an energy bin, they found that, for a background spectrum
$d\Phi/dE \sim 1/E^\alpha$, the upper limit
to the cross section scales as
$\langle \sigma v  \rangle_{limit} \sim m_{\psi_{DM}}^{3-\alpha} \Delta(\ln E)$,
$\Delta(\ln E)$ being the logarithmic energy bin. We checked that this constraint
is respected for each point of our parameter space. It is important to emphasize
that in the annulus region our result are independent of the halo profile because
the lower boundary of this region is $15^\mathrm{o}$ from the Galactic Center,
and the line of sight integral does not probe the cuspy region. Indeed, an
isothermal-cored profile gives roughly the same result in this case.

We want to stress that because of the $\sim 10\%$ energy resolution of the FERMI
telescope for 68\% containment radius, the distinction of the two lines will be difficult
for $m_{DM}\gsim 170$ GeV \cite{GLAST2}. However, the distinction with other
models with distinctive monochromatic $\gamma-$ray lines would still be possible\footnote{
We would like to thank A. Morselli for drawing our attention to this fact.}.
In the chiral square model, a third line from $\gamma-KK$ excitation final state would
be observable \cite{Chiralsquare}, whereas the $\gamma \gamma$ and $Z \gamma$ lines
in the IHDM should be centered below $m_Z$, and thus distinguishable by FERMI \cite{IHDMline}.
The $\gamma-$spectrum generated from 3 body final states \cite{Barger, Gammaradiation}
would be distinguishable from the $Z-\gamma$ line, whereas lines generated by a neutralino
DM should induce a large diffuse spectrum easily observable by FERMI.
To complete the analysis, we also checked that a $X-Z$ kinematic mixing respecting
electroweak precision test \cite{Erler} does not affect the
$\gamma-$ray line signal, as was already discussed in \cite{Us}.


We may also calculate the WIMP-nuclei elastic scattering cross section $\sigma_{\psi p}$
relevant for the direct detection of dark matter through its interaction
with nuclei in a large detector.
The points respecting WMAP constraint would give very low values for
$\sigma_{\psi p}$, below the sensitivity
of any future experiments. This comes from the fact that
the only diagrams contributing to the scattering is the t-channel $X$ exchange.
As $X$ is not directly coupled to the quarks, but couples through intermediate virtual Z-exchange,
the amplitude is highly loop suppressed as we found in preliminary results.
See \cite{Belanger:2008gy} for a comparative study of other extensions of the SM.

\subsection{comparison with LHC}

The Large Hadron Collider phenomenology of couplings generated by anomalous extra $U_X(1)$
 has recently been studied in the framework of the Green-Schwarz mechanism
 \cite{KumarWells} and higher dimensional operators \cite{Wellsteam}.
The former computed the production cross-section
of the $X$ boson from vector boson fusion at the LHC. It was shown that,
for the mass range ($M_X \sim 500-1000$ GeV), LHC could
detect the new physics through $pp \rightarrow X \rightarrow ZZ \rightarrow 4l$
processes provided $\Lambda_X \sim 100-150$ GeV. In the case of $\gamma-$ray detection
we showed that in the same framework, the FERMI satellite will be much more efficient
and will be able to probe a scale $\Lambda_X \sim 100-4000$ GeV.
The main reason is that the production of the $X$ boson occurs through vector boson
fusion and the $qqX$ coupling is suppressed by a factor $\sim g_X/\Lambda_X^2$, whereas in the
case of DM annihilation, the $\psi_{DM} \psi_{DM}X$ coupling is directly proportional
to $g_X$.


\section{Discussion and conclusion}

We have studied the phenomenology of a $U_X(1)$ extension of the Standard Model
where the SM particles
are not charged under the new abelian group and where the
Green-Schwarz mechanism ensures that the model is anomaly free.
We showed that the dark gauge field $X$, even though difficult to produce
at the LHC has however a clear signature in gamma-ray telescopes.
 This $U_X(1)$ extension has the unique feature that it generates a monochromatic
 $\gamma-$ray line from DM annihilation into $Z \gamma$.
 This is quite different from other models which predicts two lines \cite{IHDMline}
 or three lines \cite{Chiralsquare}, or one line radiated by fermions
 \cite{Gammaradiation}. Such a signature would be a smoking gun signal for
 these types of constructions\footnote{One should notice that in  a specific context with
 a decaying gravitino DM, there can exist an observable $\gamma-$ray line well below
 $M_{W^+}$ \cite{Decay}} if the finite energy resolution of FERMI allows to separate
 the different lines ($m_{DM} \lsim 170$ GeV depending on the incident angle).
 We investigated the scales reachable by the FERMI/GLAST telescope after
 5 years of running and showed the a scale $\Lambda_X \sim 1.5$ TeV
 is easily observable.
 This scale can be interpreted as the string scale where the Green-Schwarz mechanism
 cancels the anomalies of the model.
 Alternatively, the absence of any signals would
 restrict the scale of such models to lie above $\gsim 500$ GeV.
 Such sensitivity is highly competitive with the LHC.


\section*{Acknowledgments}{
The author want to thank particularly E. Dudas, S. Abel, P. Ko, A. Morselli
and A. Romagnoni for very useful discussions.  He is also grateful to
G. Belanger and S. Pukhov for substantial help with micrOMEGAs.
Likewise, the author
would like to thank the French ANR project PHYS@COL\&COS for financial support.
The work is also sponsored by the hepTOOLS Research Training Network
MRTN-CT-2006-035505,
and the European Union under the RTN programs
MRTN-CT-2004-503369.

}



\end{document}